\documentclass[5p]{elsarticle}
\journal{Astronomy and Computing}
\pdfoutput=1 

\usepackage[T1]{fontenc} 
\usepackage{acronym}
\usepackage[usenames,dvipsnames]{xcolor}
\usepackage{soul}
\usepackage{hyperref}
\usepackage{rotating}
\usepackage{multirow}
\usepackage{amssymb}
\usepackage{amsmath}
\usepackage{cleveref}
\usepackage{enumitem}

\DeclareMathOperator*{\argmin}{arg\,min}

\acrodef{ML}{machine learning}
\acrodef{DR2}{data release 2}
\acrodef{DR3}{data release 3}
\acrodef{MW}{Milky Way}
\acrodef{DM}{Dark matter}
\acrodef{LSR}{local standard of rest}
\acrodef{AHF}{Amiga Halo Finder}
\acrodef{WIMPs}{Weakly Interacting Massive Particles}
\acrodef{GC}{Galactic center}
\acrodefplural{LSR}[LSRs]{local standards of rest}
\newcommand{\mathsc}[1]{{\normalfont\textsc{#1}}}
\newcommand{\aap}{{Astron.~Astrophys.}}

\newcommand{\apj}{{The~Astrophysical~Journal}}
\newcommand{\apjl}{{The~Astrophysical~Journal~Letters}}

\newcommand{\mnras}{{MNRAS}}
\newcommand{\nat}{{Nature}}
\newcommand{\jcap}{{JCAP}}

\newcommand{\apjs}{{The~Astrophysical~Journal~Supp.~Series}}

\newcommand{\prd}{{Phys.~Rev.~D}}
\newcommand{\prl}{{Phys.~Rev.~Letters}}

\begin{document}

\begin{frontmatter}
\title{Sensitivity Estimation for Dark Matter Subhalos in Synthetic Gaia DR2 using Deep Learning}

\author[a]{A. Bazarov}

\author[a,b]{M. Benito}
\ead{mariabenitocst@gmail.com}

\author[a]{G. H\"utsi}

\author[b]{R. Kipper}

\author[a]{J. Pata}
\cortext[cor1]{Corresponding author}
\ead{joosep.pata@cern.ch}

\author[a]{S. Põder\corref{cor1}}
\ead{sven.poder@kbfi.ee}

\address[a]{NICPB, R\"avala 10, Tallinn 10143, Estonia}
\address[b]{Tartu Observatory, University of Tartu, Observatooriumi 1, T\~oravere 61602, Estonia}

\begin{abstract}
The abundance of dark matter subhalos orbiting a host galaxy is a generic prediction of the cosmological framework, and is a promising way to constrain the nature of dark matter.
In this paper, we investigate the use of machine learning-based tools to quantify the magnitude of phase-space perturbations caused by the passage of dark matter subhalos.
A simple binary classifier and an anomaly detection model are proposed to estimate if stars or star particles close to dark matter subhalos are statistically detectable in simulations.
The simulated datasets are three Milky Way-like galaxies and nine synthetic Gaia DR2 surveys derived from these.
Firstly, we find that the anomaly detection algorithm, trained on a simulated galaxy with full 6D kinematic observables and applied on another galaxy, is nontrivially sensitive to the dark matter subhalo population.
On the other hand, the classification-based approach is not sufficiently sensitive due to the extremely low statistics of signal stars for supervised training. Finally, the sensitivity of both algorithms in the Gaia-like surveys is negligible.
The enormous size of the Gaia dataset motivates the further development of scalable and accurate data analysis methods that could be used to select potential regions of interest for dark matter searches to ultimately constrain the Milky Way's subhalo mass function, as well as simulations where to study the sensitivity of such methods under different signal hypotheses. 
\end{abstract}


\end{frontmatter}

\flushbottom

\section{Introduction}
\label{sec:intro}

\ac{DM} represents roughly 84\% of the matter content in the Universe \citep{2020A&A...641A...6P}. However, unveiling its nature has proven a difficult endeavour, and none of the proposed candidates (from several extensions of the Standard Model to primordial black holes) have yet been detected.  
Cold \ac{DM} is expected to form subhalos with masses many orders of magnitude below $10^{8}\,\rm M_{\odot}$ \citep{1984Natur.311..517B}, which is roughly the mass above which galaxies can form \citep{Kitayama:2005sm, 2006MNRAS.371..885R}. 
The abundance of subhalos is dependent on the nature of \ac{DM}. This dependency can be explained, on the one hand, by the effect of the properties of the DM on the linear matter power spectrum. If for cold \ac{DM} the minimum halo mass might be as small as $10^{-12}\,\rm M_{\odot}$ \citep{1999PhLA..260..262Z, 2009NJPh...11j5027B}, microscopic properties of the DM particle, e.g. non-negligible thermal velocities or quantum pressure, introduce a cut-off at the small scales in alternative DM scenarios. On the other hand, the nature of \ac{DM}, e.g. thermal velocities or self-interactions, further impacts the non-linear growth of structures \citep{2012MNRAS.424..684S, 2016MNRAS.460.1399V}. Detecting a dark subhalo would be the first direct evidence of DM clustering at small scales. Furthermore, constraints on the subhalo abundance would provide valuable information about the particle nature of DM. 

\begin{figure*}[h]
    \centering
    \includegraphics[width=0.9\textwidth]{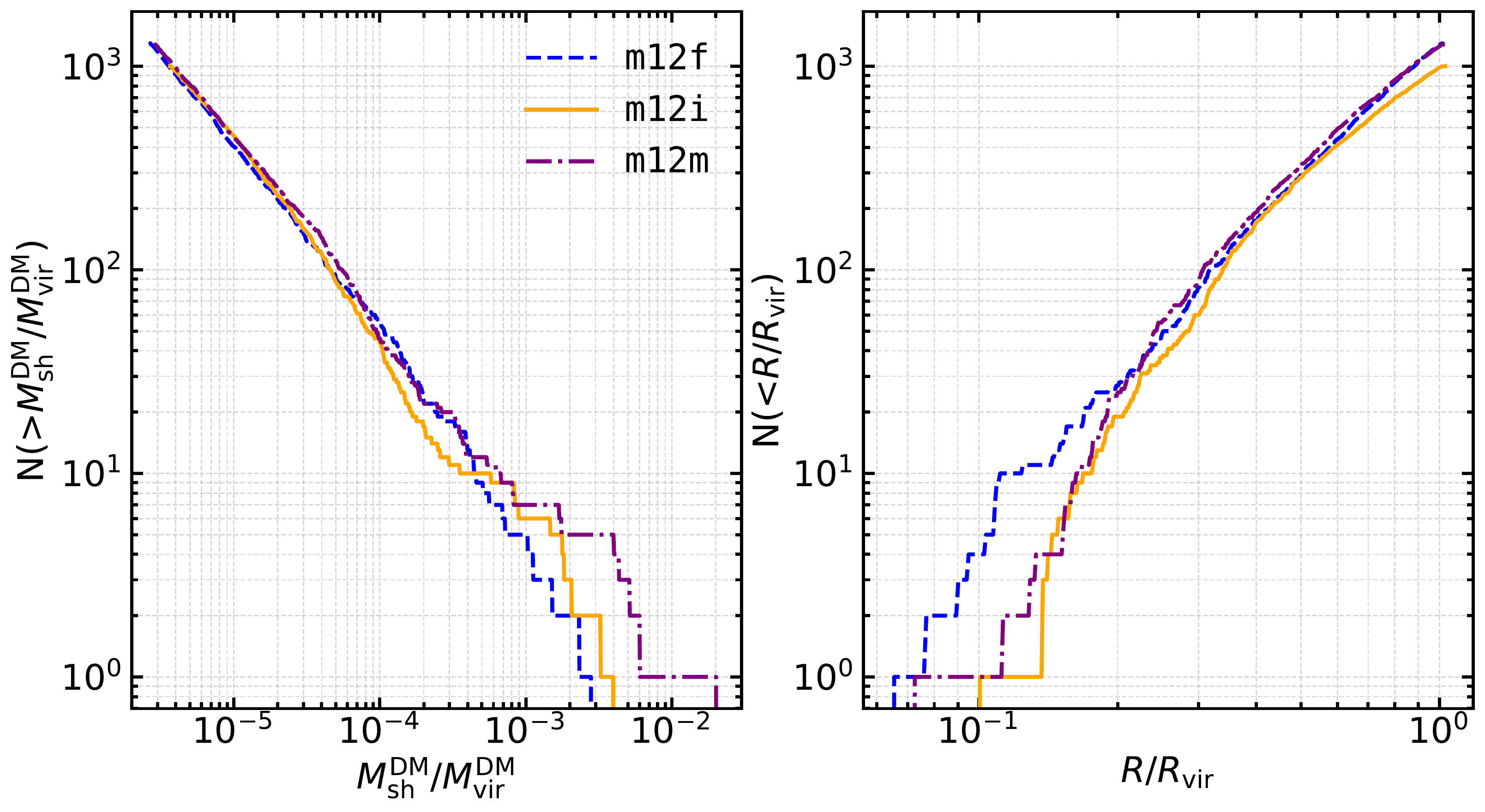}
    \caption{The subhalo mass function scaled to the host halo virial mass (left) and the radial distribution of the subhalo population (right) for the three MW-like galaxies used in this work.}
    \label{fig:subhalos_mass_radial_dist}
\end{figure*}
Subhalos with masses lower than $10^8\,\rm M_{\odot}$ are unable to form stars and remain dark \citep{Kitayama:2005sm, 2006MNRAS.371..885R}, thus hindering their detection. Strategies that aim to detect dark subhalos rely on measuring their gravitational signatures via stellar dynamics \citep{2002MNRAS.332..915I, 2011ApJ...731...58Y, 2012ApJ...748...20C, 2017MNRAS.466..628B, 2018JCAP...07..061B, 2019ApJ...880...38B, 2020PhRvD.101j3023B, 2015MNRAS.446.1000F, 2018PhRvL.120u1101B}, gravitational lensing \citep{2016JCAP...11..048H, 2018JCAP...07..041V, 2018PhRvD..98j3517D,
2019MNRAS.487.5721G,
2019ApJ...886...49B, 2020arXiv200811577V} or pulsar timing \citep{2007MNRAS.382..879S, 2011PhRvD..84d3511B, 2016MNRAS.456.1394C, 2018arXiv180107847K, 2022PhRvD.105l3514D}
and, in the case of several DM candidates, e.g. \ac{WIMPs}, on detecting the flux of final stable particles produced by \ac{DM} annihilation or decay (e.g. \citep{2010PhRvD..82f3501B, 2012ApJ...747..121A, 2012ZechlinHorns, 2017MNRAS.466.4974M, 2019JCAP...07..020C, 2019Galax...7...90C, 2019JCAP...11..045C, 2021PDU....3200845C, 2021JCAP...11..033M}). The goal of searches based on stellar dynamics is to detect perturbations in the phase-space distribution of \ac{MW} stars induced by gravitational effects of passing subhalos. We can look for these perturbations in stellar streams \citep{2002MNRAS.332..915I, 2011ApJ...731...58Y, 2012ApJ...748...20C, 2017MNRAS.466..628B, 2018JCAP...07..061B, 2019ApJ...880...38B} and in the disk or the halo stars \citep{2015MNRAS.446.1000F, 2018PhRvL.120u1101B}. In the present work we investigate the usage of an anomaly detection and classification algorithms in the search for the imprint caused in halo stars by passing substructures. In this way, we exploit the increasing size of observational datasets and state-of-the-art techniques in deep learning.

In recent years, deep learning techniques have been applied in the search for substructures in our Galaxy \citep{2020A&A...636A..75O, 2020NatAs...4.1078N, 2021MNRAS.tmp.3041S}. These detection methods assume that stars in the \ac{MW} sharing a common origin should cluster in orbital properties and/or composition. Our search differs in that we aim to identify stars that,  regardless of their origin, have their distribution in phase-space perturbed by the passage of a dark matter subhalo. For any identified star, it must be possible to test the halo hypothesis independently of the methodology used to select the candidates.
One possibility could be to preselect the stars using a \acs{ML}-based classifier, followed by detailed hypothesis tests using e.g. the orbital arc method~\citep{kipper2020quantifying, OAM2} or the stellar wakes technique~\citep{2018PhRvL.120u1101B}.

The raw data that we used, which are described in  section~\ref{sec:data}, are three \ac{MW}-like galaxies from the Latte suite of FIRE-2 simulations~\citep{2016ApJ...827L..23W} and nine synthetic Gaia DR2 surveys generated from the simulated galaxies by means of the Ananke framework~\citep{2020ApJS..246....6S}. First, we processed the synthetic Gaia datasets to correlate the position of stars and the dark subhalos, which were previously identified in the simulated galaxies.
In section~\ref{sec:detectability}, we estimate the detectability of the subhalo-associated stars using deep learning techniques.
We conclude in section~\ref{sec:conclusions}.

\section{Datasets}
\label{sec:data}

As our raw data, we used three \ac{MW}-like galaxies from the Latte suite of FIRE-2 simulations~\citep{2016ApJ...827L..23W, garrison2017not, hopkins2018fire} (dubbed {\tt m12f}, {\tt m12i} and {\tt m12m}) and nine synthetic Gaia DR2 surveys~\citep{2020ApJS..246....6S}. This section describes these datasets and the processing we performed on them.

\begin{figure*}[h]
    \centering
    \includegraphics[width=0.95\textwidth]{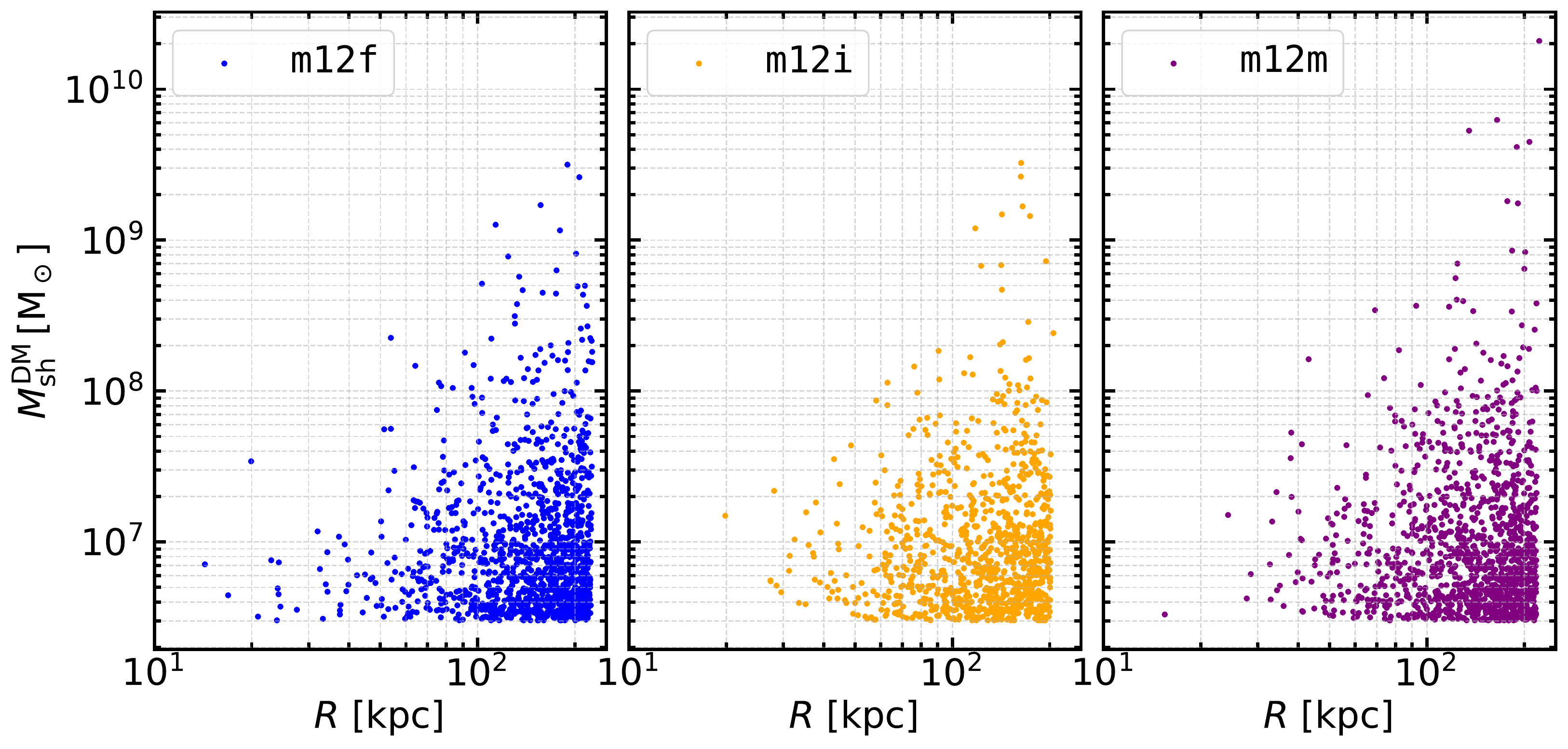}
    \caption{Subhalo mass as a function of galactocentric distance. Each dot corresponds to a subhalo as identified by \ac{AHF} for {\tt m12f} (left), {\tt m12i} (middle) and {\tt m12m} (right) galaxies.}
    \label{fig:subhalos_mass_vs_radius}
\end{figure*}
\subsection{Milky Way-like Galaxies}
\label{subsec:Nbody}
We used the simulation snapshots at $z=0$ of three MW-like galaxies\footnote{Taken from\\ \href{https://girder.hub.yt/\#collection/5b0427b2e9914800018237da}{https://girder.hub.yt/\#collection/5b0427b2e9914800018237da}}, namely  {\tt m12f}, {\tt m12i} and {\tt m12m}~\citep{2016ApJ...827L..23W, garrison2017not, hopkins2018fire}. In the following we briefly describe how these MW analogues were obtained. For a complete description of this and the details of the N-body simulations we refer the interested reader to~\citep{2016ApJ...827L..23W} and references therein.
The MW analogues were first identified in a DM-only cosmological simulation requiring that at $z=0$: (i) their virial mass is in the range of $M_{200}=[1-2]\times 10^{12}\,\rm M_\odot$\footnote{Virial mass and virial radius follow the relation $M_{200}=\frac{4\pi}{3}200\rho_m R_{200}^3$, with $\rho_m$ the average matter density of the Universe.} (which agrees with recent measurements \citep{2020SCPMA..6309801W, 2020JCAP...05..033K, 2021arXiv211109327S}) and (ii) there is no neighboring halo of similar mass within $5R_{200}$. Three halos selected in this manner were then simulated using the zoom-in technique~\citep{2014MNRAS.437.1894O}. 
Simulations were run using the Gizmo gravity plus hydrodynamics code in meshless finite-mass (MFM) mode~\citep{2015MNRAS.450...53H} and the FIRE-2 baryonic physics model~\citep{hopkins2018fire}. Dark matter particles in the zoom-in simulation have a mass of $m_{\rm DM}=3.5\times10^{4}\,\rm M_\odot$, and the initial gas or star particle mass is $m_{\rm gas}=7.1\times 10^3\,\rm M_\odot$. Dark matter and stars have gravitational softening lengths $\epsilon_{\rm DM}=20\,\rm pc$ and $\epsilon_{\rm star}=4\,\rm pc$, respectively. The softening lenght of the gas is adaptive, and reaches a minimum value of $\epsilon_{\rm gas, min}=1\,\rm pc$.

We identified \ac{DM} subhalos in snapshots at $z=0$ of the MW-like galaxies using the Amiga  Halo Finder (\ac{AHF})code \citep{knollmann2009ahf}.
The \ac{AHF} algorithm identifies bound \ac{DM} structures by hierarchically clustering 3D positions of \ac{DM} particles in the simulation. Following~\citep{garrison2017not}, \ac{AHF} was run only on DM particles. We selected subhalos with more than 85 \ac{DM} particles
(corresponding to subhalos with masses $M_{\rm sh}^{\rm DM}>3\times 10^6\,\rm M_\odot$) since those substructures are reliably resolved in the simulation~\citep{garrison2017not}. However, the MW is expected to have a population of subhalos with lower masses. The velocity changes in stars due to the gravitational encounter with a dark subhalo with a mass of $10^5\,\rm M_\odot$ are of the order of $10^{-3}\,\rm km/s$ \citep{2015MNRAS.446.1000F}, which are well below the statistical uncertainties in observations of MW halo stars. Therefore, we argue that subhalos with masses smaller than $3\times10^{6}\,\rm M_\odot$ have a negligible impact on the current investigation of the feasibility of two simple algorithms that search for perturbations in each star independently of each other. 
Nonetheless, we leave a thorough investigation of the detection of subhalos unresolved in the simulation for a follow-up work.

Approximately $\simeq10^3$ subhalos\footnote{\ac{AHF} identifies 1298, 1001 and 1281 subhalos with $N_{\rm DM}>85$ for {\tt m12f}, {\tt m12i} and {\tt m12m}, respectively.} for each \ac{MW}-like galaxy remain as potentially observable. Figure~\ref{fig:subhalos_mass_radial_dist} shows the cumulative subhalo mass function normalized by the virial mass of the host halo (left panel) and the radial distribution of the subhalo population normalized by the virial radius (right panel). The virial masses are $M_{\rm vir}^{\rm DM}=1.1\times10^{12}\,\rm M_{\odot}$, $0.8\times10^{12}\,\rm M_{\odot}$ and $1.0\times10^{12}\,\rm M_{\odot}$ for {\tt m12f}, {\tt m12i} and {\tt m12m}, respectively.\footnote{In our work we have used the virial mass definition given by $M_{\rm vir}=\frac{4\pi}{3} 178 \rho_{\rm crit} R_{\rm vir}^3$, with $\rho_{\rm crit}$ the critical density of the Universe at $z=0$.}
In figure \ref{fig:subhalos_mass_vs_radius} we show the mass of the subhalos as a function of their galactocentric distance for {\tt m12f}, {\tt m12i} and {\tt m12m}. It should be noted that no subhalos are identified below $14\,\rm kpc$ from the center of the galaxies, as previously noted in~\citep{garrison2017not}. Furthermore, 97\%, 91\% and 94\% of the subhalos below 50 kpc for {\tt m12f}, {\tt m12i} and {\tt m12m}, respectively, have masses lower than $1\times10^{7}\,\rm M_\odot$. The most massive subhalo below 50 kpc is identified at 20 kpc with $M_{\rm sh}^{\rm DM}=3\times10^{7}\,\rm M_\odot$ for {\tt m12f}, at 43 kpc with $M_{\rm sh}^{\rm DM}=4\times10^{7}\,\rm M_\odot$ for {\tt m12i} and at 43 kpc with $M_{\rm sh}^{\rm DM}=2\times10^{8}\,\rm M_\odot$ for {\tt m12m}. The depletion of the most massive dark subhalos in the inner 50 kpc of the MW-like galaxies conditions the ability to identify stars in the stellar halo which have been perturbed by the passage of a dark matter subhalo using the deep-learning techniques explored in this study.

\subsection{Synthetic Gaia Surveys}
\label{subsec:synthGaia}
The nine synthetic Gaia DR2 surveys were generated by applying the Ananke framework~\citep{2020ApJS..246....6S} to the three MW-like galaxies. Per simulated galaxy, three synthetic surveys were generated by adopting three \acp{LSR}. Each synthetic survey contains approximately a billion mock stellar observations resembling Gaia DR2. 
We restrict our attention to stellar halo stars, applying a selection in true vertical distances $|z| > 5$ kpc. 
In this way, we remove disk stars that could suffer from disturbances induced, for example, by spiral arms, the Galactic bar or giant molecular clouds. We would like to highlight that the disk was primarily excluded because the data volume was too large to cope with at this stage. 
Notice that this problem, however, does not affect the MW-like simulations where the number of star particles per galaxy is of the order of $10^7$. It is not clear how our results would be affected if we were to include the disk in our analysis of the synthetic survey, and it is a question that we plan to address in a follow-up. After removing the disk, we are left with $\mathsc{O}(10^{8})$ mock stars for each \ac{LSR} for the subsequent analysis. This reduced dataset consists of nearly 2 billion observed stars for the three different \ac{MW}-like galaxies, three \acp{LSR} for each, correlated with potentially observable \ac{DM} subhalo locations. 
Table~\ref{tab:dataset} summarizes some statistics of this dataset.

\renewcommand{\arraystretch}{1.4}
\begin{table*}[h] 
\centering
\small\addtolength{\tabcolsep}{3pt}
\resizebox{\textwidth}{!}{\begin{tabular}{ | c | c || c |  c || c | c || c |}
\hline
\multicolumn{2}{|c||}{}  & stars with $|z|>5$ kpc & with $v_r$ [$\%$] & halo-associated stars [$\%$] & halo-associated stars with $v_r$ [$\%$] &  subhalos with associated stars \\
\hline \hline
\multirow{3}{*}{{\tt m12f}}     & LSR0 & 216,446,024 & 0.42\% & 0.0291\% & 0.35\% & 73 \\
\cline{2-6} 
                                & LSR1 & 182,538,592 & 0.44\% & 0.0291\% & 0.32\% & 76 \\
\cline{2-6} 
                                & LSR2 & 204,017,261 & 0.44\% & 0.0306\% & 0.35\% & 71 \\
\hline \hline
\multirow{3}{*}{{\tt m12i}}     & LSR0 & 139,167,343 & 0.45\% & 0.0019\% & 0.41\% & 63 \\
\cline{2-6} 
                                & LSR1 & 132,655,442 & 0.46\% & 0.0017\% & 0.41\% & 61 \\
\cline{2-6} 
                                & LSR2 & 131,474,668 & 0.48\% & 0.0010\% & 0.23\% & 67 \\
\hline \hline
\multirow{3}{*}{{\tt m12m}}     & LSR0 & 170,255,144 & 0.47\% & 0.0013\% & 0.09\% & 67 \\
\cline{2-6} 
                                & LSR1 & 156,093,757 & 0.47\% & 0.0016\% & 0.12\% & 71 \\
\cline{2-6} 
                                & LSR2 & 161,369,511 & 0.47\% & 0.0013\% & 0.19\% & 68 \\
\hline
\end{tabular}}
\caption{\label{tab:dataset}
Summary statistics of synthetic Gaia DR2 reduced catalogs used in this work (see text for more details).}
\end{table*}

\begin{figure*}[h]
    \centering
        \includegraphics[width=0.9\textwidth]{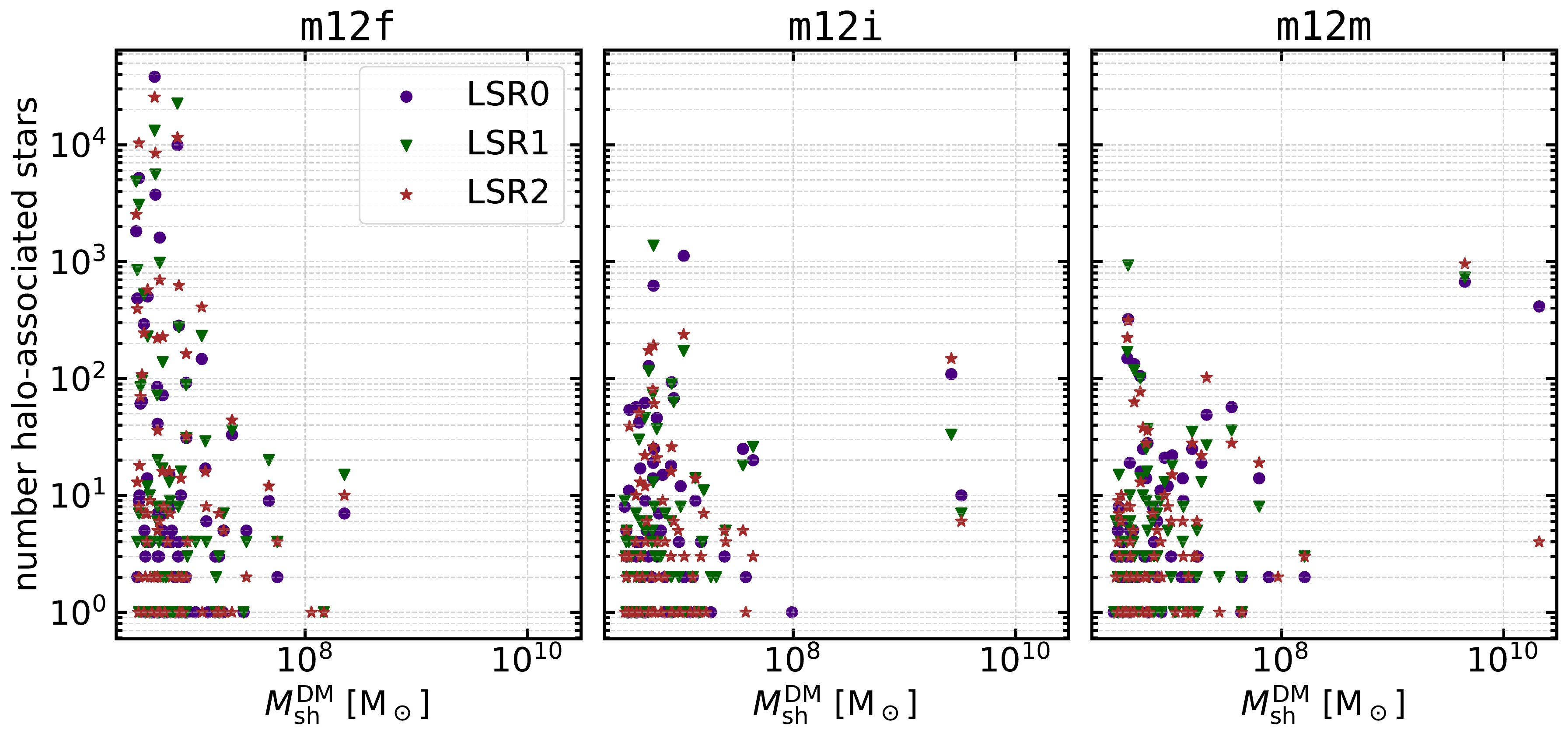}
    \caption{Number of stars associated to a particular subhalo as a function of its mass for {\tt m12f} (left), {\tt m12i} (middle) and {\tt m12m} (right) galaxies.}
    \label{fig:subhalos_stars_mass}
\end{figure*}

\begin{figure*}[t]
    \centering
    \includegraphics[width=0.9\textwidth]{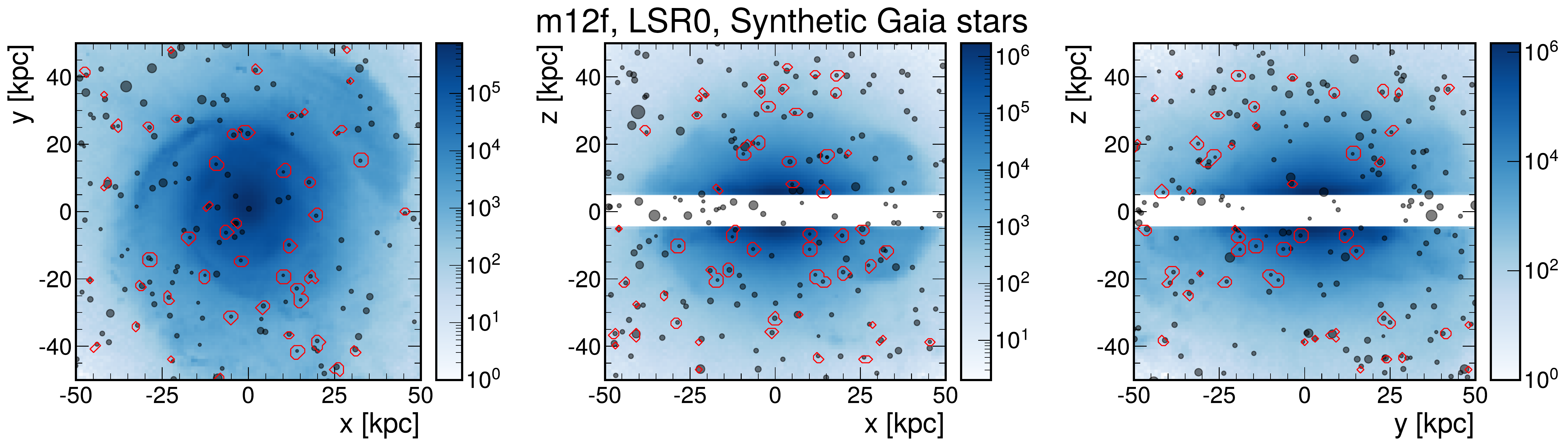}\\
    \includegraphics[width=0.9\textwidth]{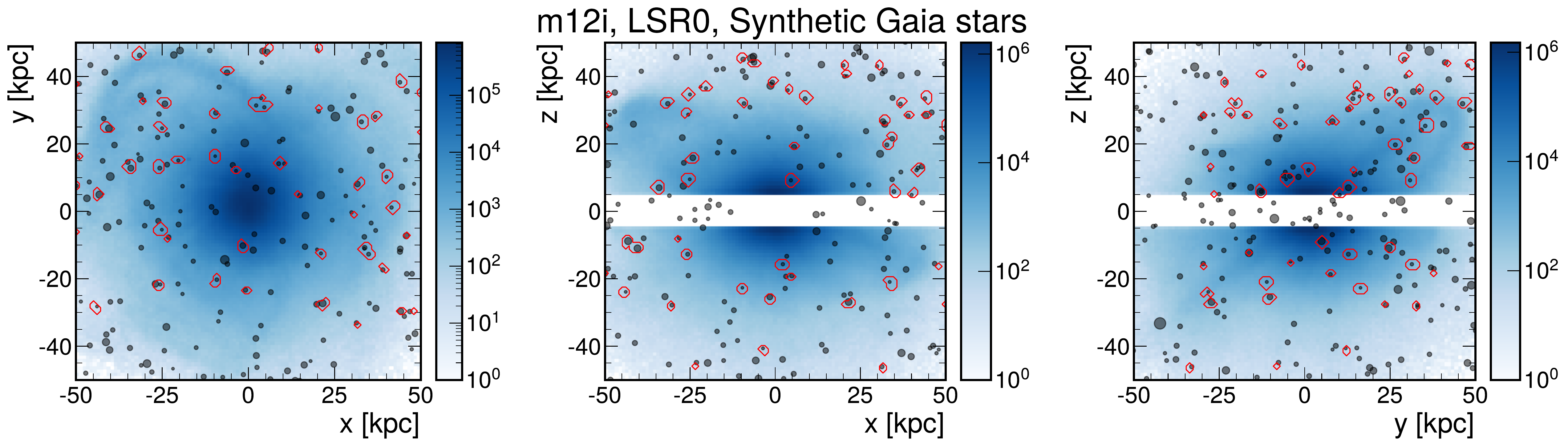}\\
    \includegraphics[width=0.9\textwidth]{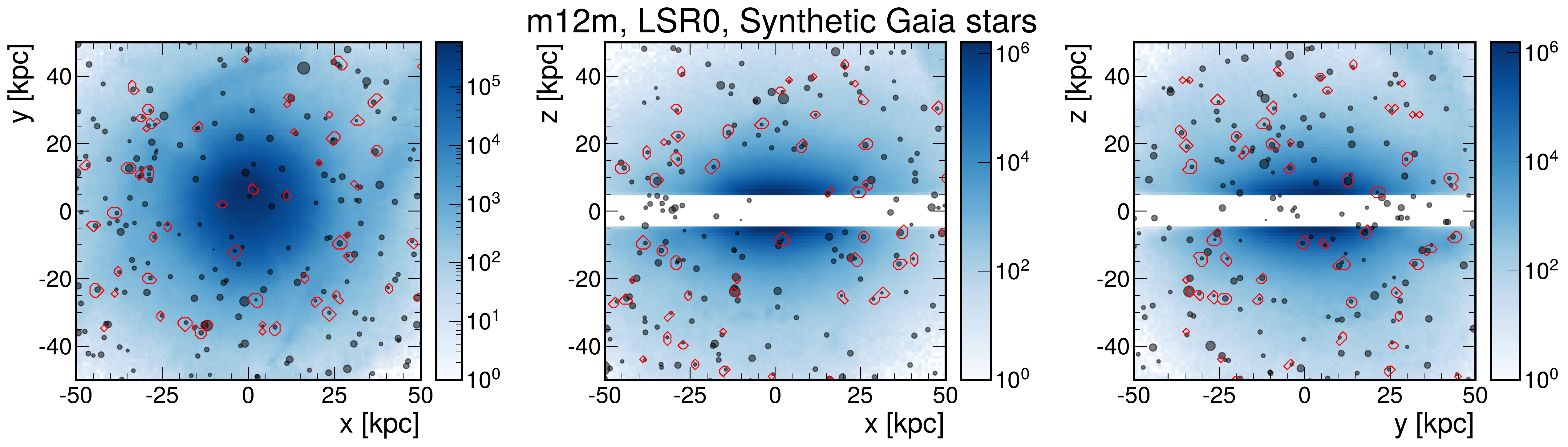}\\
    \caption{Projected stellar number densities in the synthetic Gaia datasets in true galactocentric coordinates for the three MW-like galaxies, namely {\tt m12f} (top row), {\tt m12i} (middle row) and {\tt m12m} (bottom row) for LSR0. The halo locations are shown in black, with the size of the markers being proportional to the halo's virial radius, while the regions with halo-associated observed stars are shown in red. The Sun's position for each case is $(x, y, z) = (0, 8.2\,{\rm kpc}, 0)$.}
    \label{fig:stars_gaia_lsr0}
\end{figure*}

Stars are tagged as halo-associated if their true distance to the central position of a subhalo is lower than 1 kpc. It is to be noticed that these halo-associated stars might not be bound to the subhalos (see next section).  Figure~\ref{fig:subhalos_stars_mass} shows the total number of stars associated to a subhalo as a function of the subhalo's mass for each LSR and each simulated galaxy. Within each galaxy, less than $\sim 10\%$ of the subhalos contain associated stars, and 66\%, 84\% and 75\% of this fraction contain less than 10 associated stars for {\tt m12f}, {\tt m12i} and {\tt m12m} galaxies, respectively. Furthermore, approximately 40\% of the halos that have associated stars contain only one star.
The {\tt m12f} galaxy has a larger percentage of subhalos which are associated to more than 100 stars compared to that of {\tt m12i} or {\tt m12f} galaxies. This is because the former galaxy has a larger fraction of subhalos below 30 kpc.
We plot the projected stellar number densities for LSR 0 on Figure~\ref{fig:stars_gaia_lsr0}, along with the halo locations and halo-associated observed stars.

\section{Deep Learning Search of Subhalo-associated Stars}
\label{sec:detectability}

Dark subhalos perturb the positions and velocities of nearby stars. We wish to estimate if these kinematic imprints are detectable in MW-like galaxies and in synthetic Gaia data that accounts for observational uncertainties.
Let us assume, without loss of generality, that the properties $\mathbf{X}$ of each star particle (or observed star) are drawn from the probability distribution $p(\mathbf{X} | \mathrm{sig})$ or $p(\mathbf{X} | \mathrm{bkg})$ if the star particle (observed star) has or has not been affected, respectively, by a dark subhalo at a given time in the Latte (Ananke) simulation. 
Then, if the probabilities are known, the likelihood ratio  
$p(\mathbf{X} | \mathrm{sig})/p(\mathbf{X} | \mathrm{bkg})$ is the optimal discriminator between the two hypotheses for a given observation according to the Neyman-Pearson lemma~\citep{neyman1992problem}.
These probabilities are not known, however, and we only have simulated examples of either halo-associated or background star particles (observed stars).
In the following, we investigate the possibility of using machine learning to define an approximate discriminator between the two hypotheses, and thus quantify the difference between the halo-associated stars and the background.

\subsection{Detectability}
\label{subsec:detectability}
As a starting point, we first focus on the Latte simulations, where for each star particle, the full six-dimensional phase-space coordinates, namely the three-dimensional Galactocentric Cartesian positions and velocities, are known.
Unlike for the synthetic Gaia dataset, the disk is not excluded at this stage. In addition, here we only consider subhalos with galactocentric distances less than 100 kpc. We are then left with subhalos with masses smaller than $4 \times 10^8\rm M_\odot$, which reduces the probability of including stars associated with the halo of dwarf galaxies. 
This cut in radius, or equivalently in mass, does not strictly mean that luminous halos are excluded from our catalogue. For this reason, we have identified candidate dwarf galaxies as those subhalos that have more than one signal star with a relative velocity with respect to the subhalo smaller than the subhalo's escape velocity. In this manner we have identified 4, 6 and 9 subhalos for {\tt m12f}, {\tt m12i} and {\tt m12m}, respectively. By removing these subhalos in the analyses presented in this section, our results are quantitatively the same.

For each star particle, we compute the Euclidean distance to the nearest dark subhalo $d$, and if it is below a threshold $d < d_{\mathrm{max}}=1$~kpc, we identify the star particle as a halo-associated or signal particle.
We then use an anomaly detection approach to estimate the strength of the subhalo signal~\citep{10.1016/0893-6080(89)90014-2,Sakurada2014AnomalyDU}.
For this purpose, the background-only likelihood $L_b(\mathbf{X}) \simeq p(\mathbf{X} | \mathrm{bkg})$ is indirectly approximated using a so-called autoencoder neural network, and deviations from the background-only distribution are quantified. 

Each star particle is characterized by the feature vector $\mathbf{X}$ containing its three-dimensional position and velocity, i.e. $(x, y, z, v_x, v_y, v_z)$.
Let us define an encoder $E(\mathbf{X})$ and a decoder $D(\mathbf{z})$ as
\begin{align}
    E(\mathbf{X}) &\rightarrow \mathbf{z} \in \mathbb{R}^D\textrm{ and}\\
    D(\mathbf{z}) &\rightarrow \mathbf{X}' \in \mathbb{R}^6,
\end{align}
respectively, such that $D(E(\mathbf{X})) \rightarrow \mathbf{X}'$ approximates $\mathbf{X}$ for any given input via a lower-dimensional $D<6$ representation.
Both the encoder and decoder are implemented as feedforward neural networks, optimized by tuning the weights using only the background examples as follows:
\begin{equation}
     \mathbf{D,E} = \argmin_{D,E} \sum_{i\in\mathrm{bkg}} \left\Vert\mathbf{X}_i - D(E(\mathbf{X}_i))\right\Vert.
\end{equation}
\begin{figure*}
    \centering
    \includegraphics[width=0.45\textwidth]{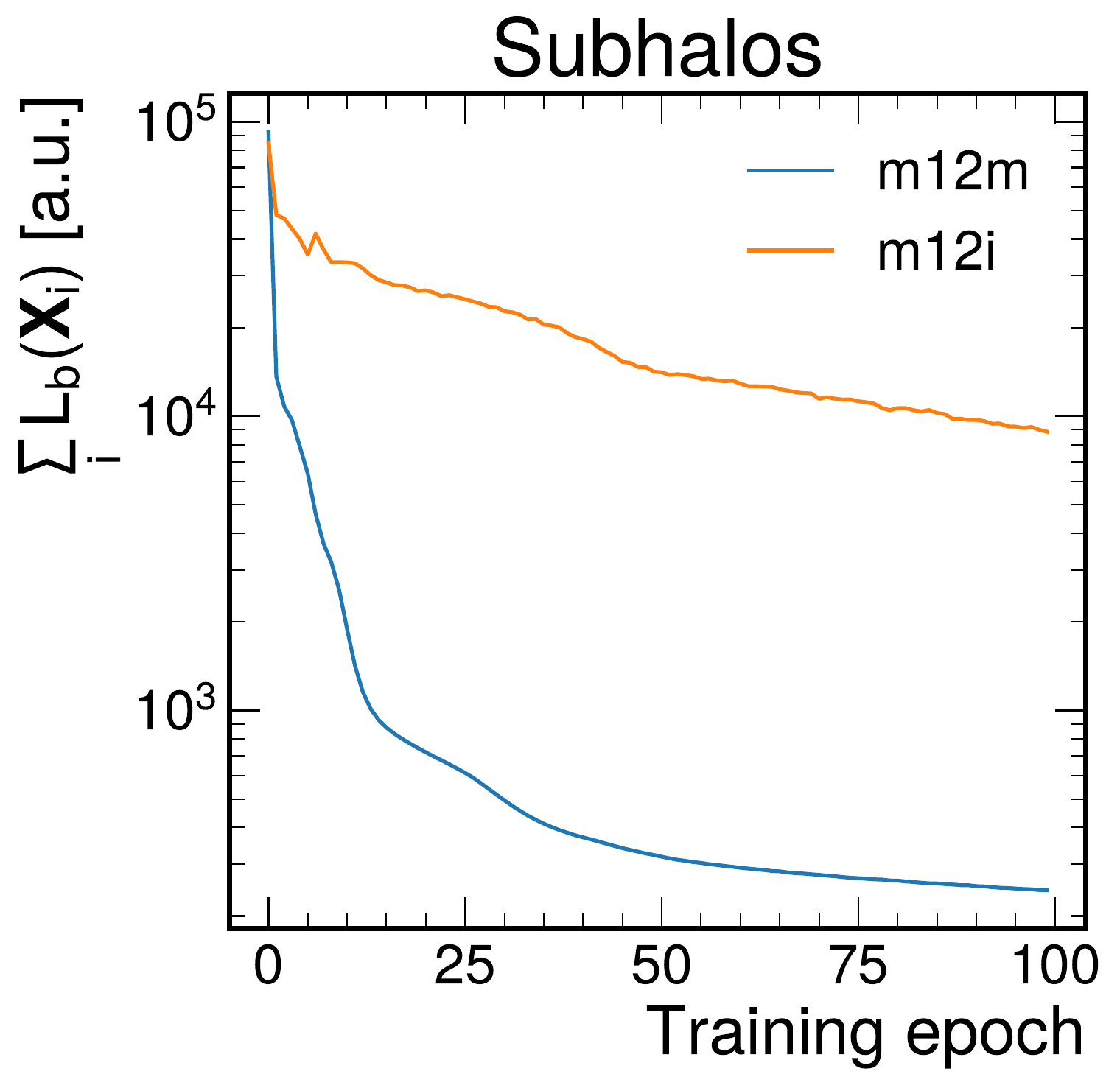}
    \includegraphics[width=0.45\textwidth]{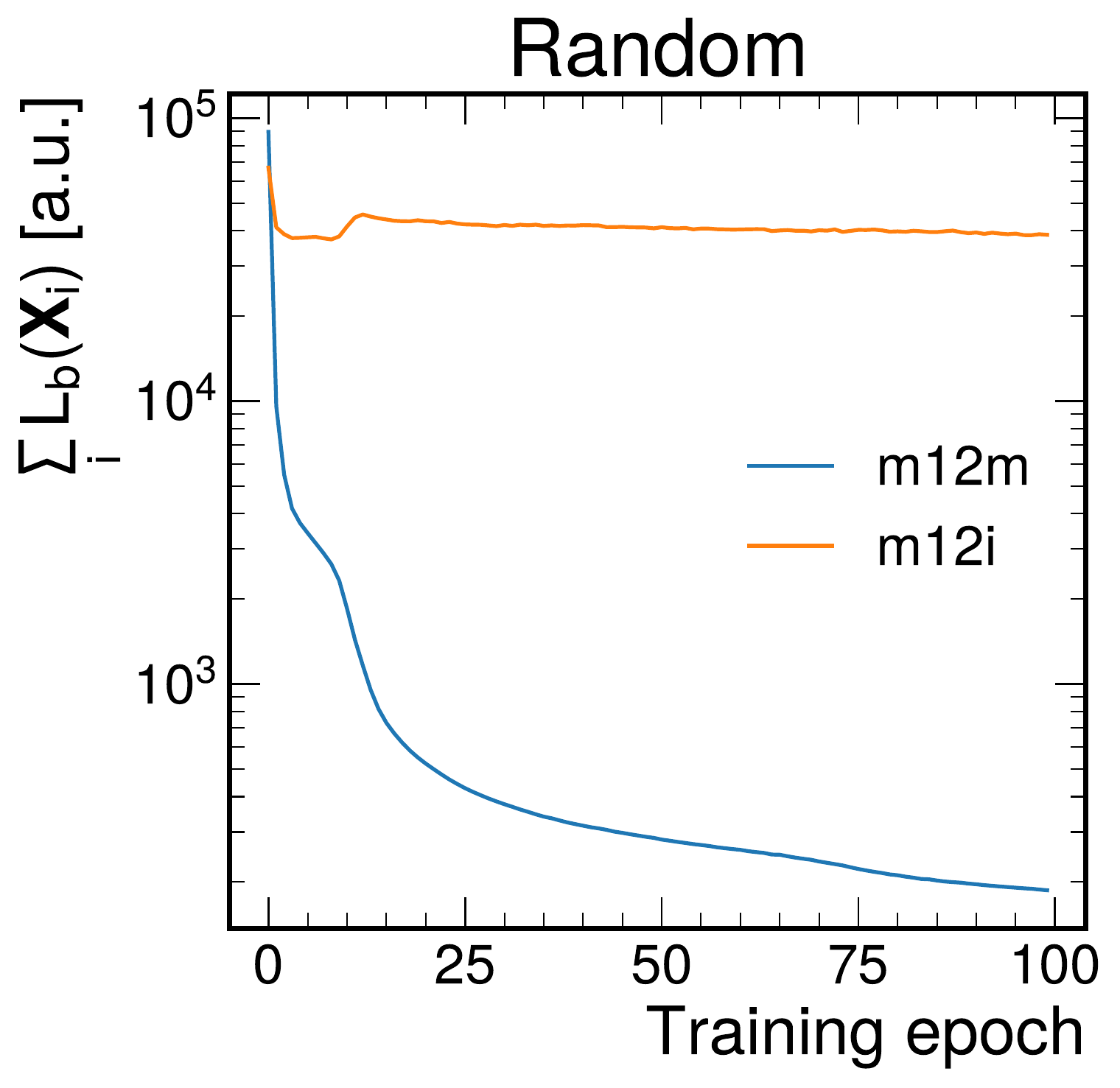}\\
    \caption{Relative reconstruction loss with respect to the first epoch for the training ({\tt m12m}) and validation ({\tt m12i}) datasets based on the anomaly detection model over the training epochs. By construction, we expect the loss in {\tt m12m} to decrease as the model fits the samples. The training is performed on star particles excluding the subhalo-associated particles (left) and by excluding the same number of random particles (right) as a check. A.u. stands for arbitrary units of the loss function. We note that the underlying signal distribution of the training and validation datasets are different, thus the numerical values of the loss functions are not necessarily directly comparable between the training and validation datasets. We observe no significant overtraining on the validation dataset.}
    \label{fig:latte_autoencoder_loss}
\end{figure*}
The neural network model parameters, such as the number of layers and neurons in each layer, the size of the lower-dimensional representation and the activation function, are chosen based on a small number of experiments rather than through a systematic hyperparameter optimization, which is left for a future study. We use two layers with $128$ neurons for the encoder, the latent space $D=3$, and two layers for the decoder, with again $128$ neurons per layer. We use the scaled exponential linear unit (SELU) activation function for the hidden layers~\citep{klambauer2017self}.

By construction, the encoder-decoder will tend to reconstruct well the background-like samples that it was optimized on.
On the other hand, for any other $\mathbf{X}$ that is not distributed as $p(\mathbf{X} | \mathrm{bkg})$, we would expect on average higher values for the reconstruction loss $L_b(\mathbf{X}_i) = \left\Vert\mathbf{X}_i - D(E(\mathbf{X}_i))\right\Vert$.
Therefore, we can use the distribution $L_b(\mathbf{X})$, optimized only on the background particles, as an empirical discriminator between the background and signal samples.
We have checked this approach by defining a fake signal consisting of a random sub-population of stars irrespective of the dark subhalo locations.
In this case, no detectable difference between the main sample and the random subpopulation of stars is expected with this method.

We optimize the model on the {\tt m12m} galaxy, while cross-checking the performance on the {\tt m12i} galaxy.
This ensures that the model is not simply memorizing the locations of the halos, as the result in this case would not be generalizable to other galaxy simulations.
Training is carried out for 100 iterations (epochs) over the full dataset using the Adam optimizer~\citep{2014arXiv1412.6980K} with a learning rate of $l=10^{-4}$ and a minibatch size of $10^5$ star particles.
We show the evolution of the total reconstruction loss over training epochs on figure~\ref{fig:latte_autoencoder_loss} for both the real subhalo signal (left panel) and the fake signal cases (right panel).
We observe that the model converges for the training dataset {\tt m12m} and exhibits in general stable behaviour for the validation dataset {\tt m12i}.

\begin{figure*}
    \centering
    \includegraphics[width=0.45\textwidth]{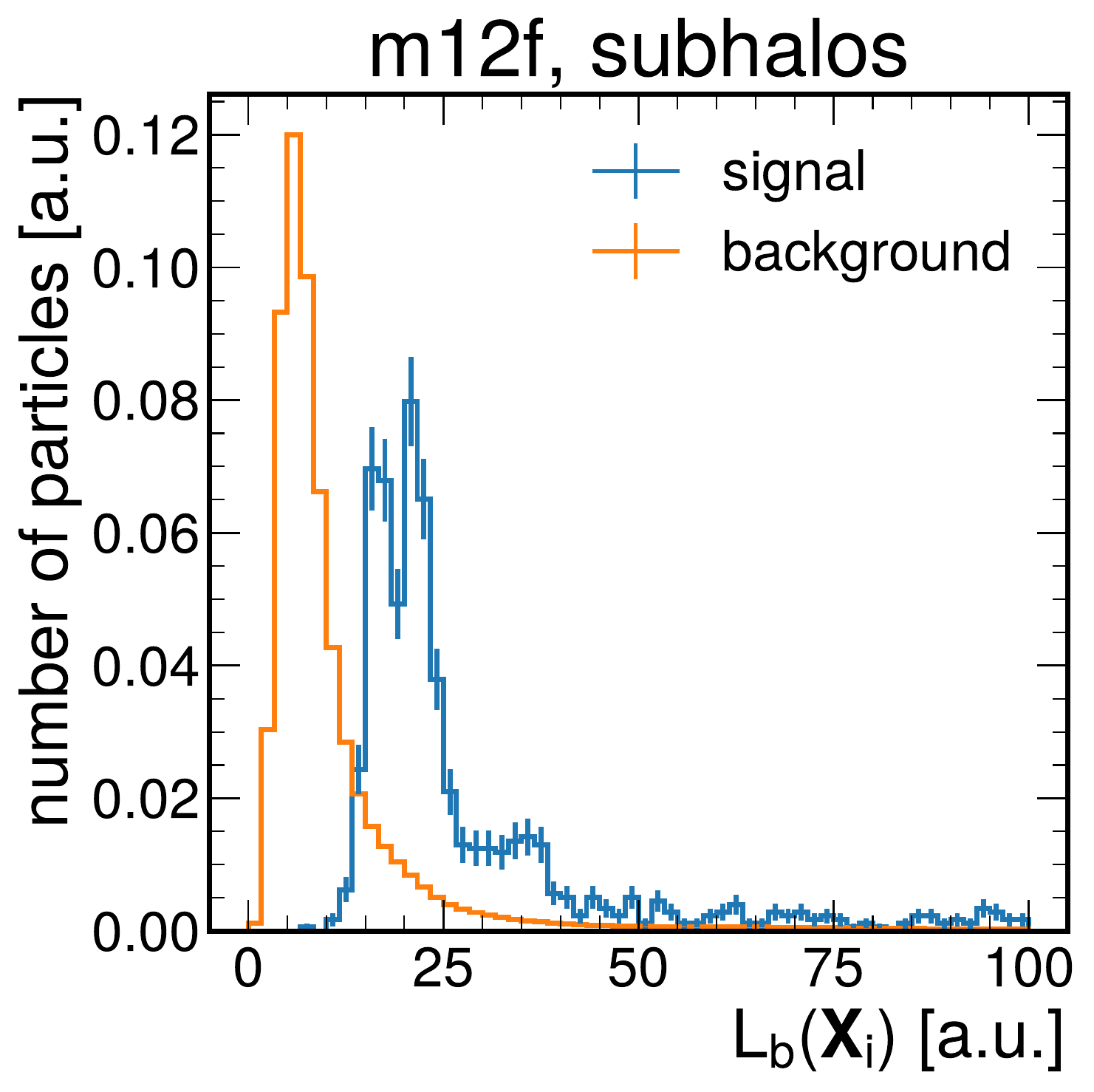}
    \includegraphics[width=0.45\textwidth]{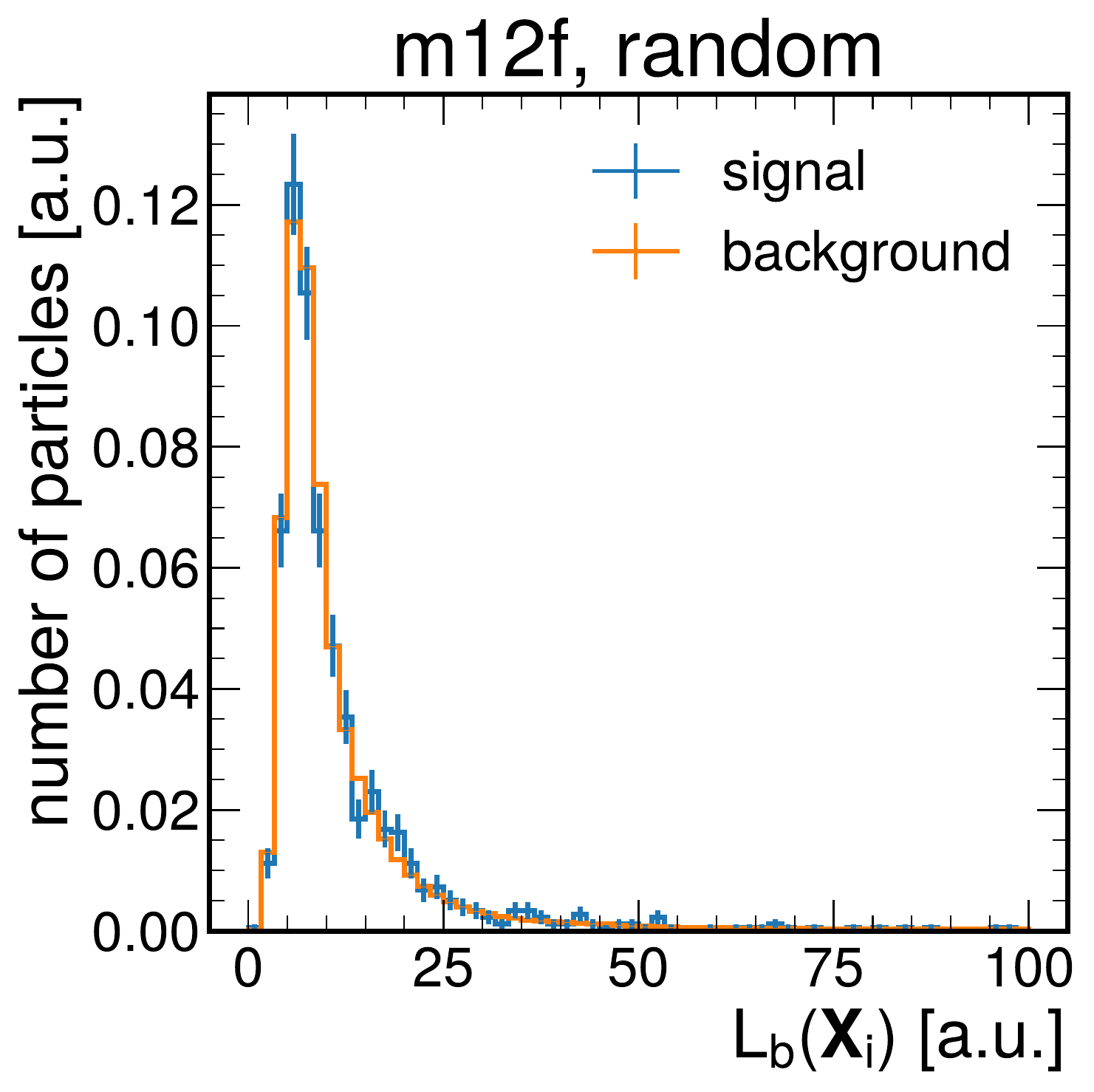}\\
    \caption{The distribution of the reconstruction loss $L_b$ for the {\tt m12f} galaxy that was never used in the training procedure. We show the distributions of the halo-associated (signal) and the rest (background) of the star particles separately. On the left, the model was trained and evaluated with the subhalo-associated as the signal, while on the right, a random set of stars was denoted as the signal as a check.}
    \label{fig:latte_autoencoder_pred}
\end{figure*}

Figure~\ref{fig:latte_autoencoder_pred} shows the $L_b(\mathbf{X}_i)$ distributions for the signal and background stars for the {\tt m12f} dataset never used for training. 
It is observed that for the real subhalo signal case (left panel), there is a distinction between the distribution of halo-associated (signal) and non-halo associated (background) stars, with the signal stars having on average higher values of the reconstructed distribution.
No such distinction is observed for the model trained and tested on the random subset (right panel), as would be expected.

\begin{figure}
    \centering
    \includegraphics[width=0.45\textwidth]{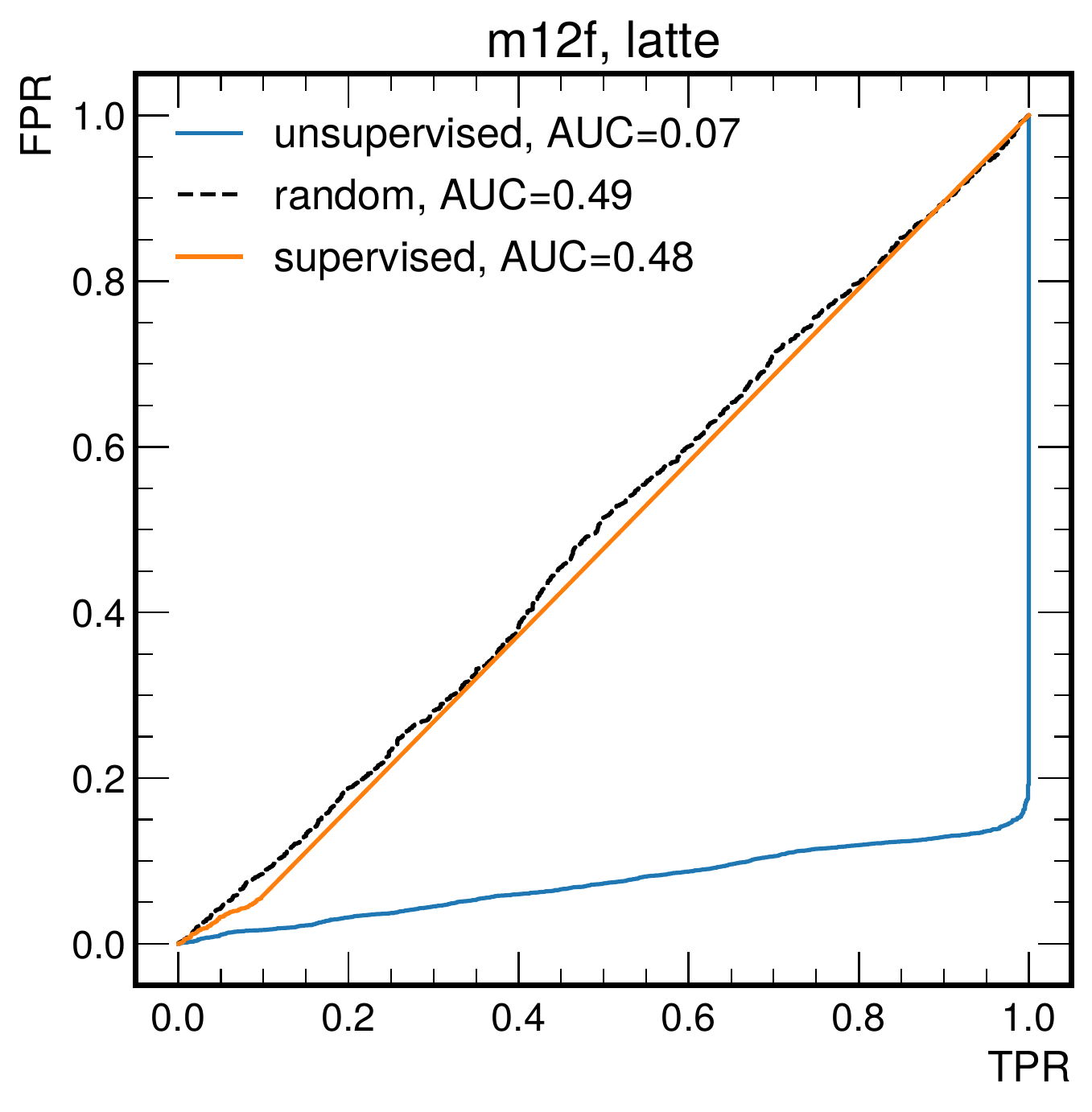}
    \caption{The true positive rate (horizontal axis) vs. the false positive rate (vertical axis). The blue line depicts the classification performance when the reconstruction loss $L_b$ is used as a discriminator between the signal and background labels on the star particles from the {\tt m12f} galaxy. The orange line shows the binary classifier performance as evaluated on the same galaxy. By construction, we observe no distinction for random stars, while the unsupervised model distinguishes halo-associated stars from the background using a combination of the position and velocity information.
    The bad performance of the supervised classifier is expected as the number of signal stars is very low compared to the background.
    }
    \label{fig:latte_autoencoder_perf}
\end{figure}

We quantify the performance of the anomaly detection in terms of the true positive and false positive rates. The true positive rate (TPR) gives the fraction of signal stars that are correctly identified as signal particles at a particular threshold $t$ (i.e. a given value of $L_b$), 
\begin{equation}
    \mathrm{TPR}(t) = \frac{N_{\mathrm{sig}}(L_b > t)}{N_{\mathrm{sig}}}.
\end{equation}
Contrary, the false positive rate (FPR) is the fraction of background stars that are  incorrectly identified as signal, namely
\begin{equation}
    \mathrm{FPR}(t) = \frac{N_{\mathrm{bkg}}(L_b > t)}{N_{\mathrm{bkg}}}.
\end{equation}
Figure~\ref{fig:latte_autoencoder_perf} shows the FPR versus TPR while scanning over $t$ for the real (solid blue) and random (dashed black) signal cases.
Using the unsupervised anomaly detection model, we see that at a $\mathrm{TPR}\simeq80\%$, the FPR is $\simeq 15\%$ (i.e. 80\% of the signal stars are correctly identified while we misclassify 15\% of the background stars as signal), presenting a significant improvement over a random selection.

We cross-check the anomaly detection approach against a simple binary classifier, where the signal model is used explicitly, but which is thus limited by the available statistics for the halo-associated stars. Contrary, in the anomaly-detection based approach, the star labels based on the proximity to a dark subhalo were only used to exclude the signal samples from the optimization. 
The supervised classification model uses the signal sample labels directly, i.e. the optimization target is the star label $y_i=\{0, 1\}$ for background and signal stars, respectively. Thus, it can be used to determine the upper limit detectability for this particular signal model, assuming training statistics are not a limiting factor.

The binary star classification model is defined as a parametric function using a deep neural network
\begin{equation}
\label{eq:cls_neural_net}
\Phi(\mathbf{X}_i | \mathbf{w}) \rightarrow y_i' \in [0,1],
\end{equation}
which can be optimized by tuning the weights $\mathbf{w}$ to minimize a classification loss function.
As before, the hyperparameters of the neural network were chosen based on a manual optimization, rather than a dedicated hyperparameter scan which is left for a subsequent study.
We use two hidden layers with 256 elements each, the SELU activation function and dropout with a coefficient of $p=0.3$. The dropout regularization~\citep{srivastava2014dropout} limits the amount of overtraining. Finally, we use the the focal loss, which is a modification of the binary cross entropy loss, originally proposed for rare object detection in~\citep{DBLP:journals/corr/abs-1708-02002}, and is defined as the following sum over the total number of star particles $N_{\mathrm{star}}$ in the dataset:
\begin{equation}
  L = \sum_{i=1}^{N_{\mathrm{star}}} -y_i \alpha(1 - y_i')^\gamma\log(y_i') - (1-y_i)(1-\alpha)y_i'^\gamma\log(1-y_i'), \\
\end{equation}
where $\alpha$ and $\gamma$ are empirical factors that adjust the weight of easy-to-classify background-like examples in the loss.
We choose $\alpha=0.25, \gamma=2.0$ based on the defaults introduced in~\citep{DBLP:journals/corr/abs-1708-02002}.
By this construction, the model output $y_i'$ for star $i$ is a continuous value between 0 and 1 that can be interpreted as a test statistic for the star being labeled as signal.

As before, we use {\tt m12m} for the optimization, 
{\tt m12i} for the validation, while {\tt m12f} is used for testing.
As shown by the orange line on figure \ref{fig:latte_autoencoder_perf}, the supervised binary classifier has a performance comparable to random selection on this dataset. The negligible sensitivity of the classifier compared to the autoencoder is to be expected due to the very low number of independent signal stars (a few thousand stars per galaxy associated with less than a hundred subhalos).

\subsection{Feasibility in Synthetic Gaia Survey}
\label{subsec:feasibility}
In this section, we investigate if the halo-associated stars are detectable in the synthetic Gaia surveys derived from the very same simulations, that is, under the effects of extinction, partial measurement of the radial velocity $v_r$ and measurement errors.
This is done by searching for dark subhalos on the reduced synthetic surveys described in section~\ref{subsec:synthGaia}. 
The goal is again to select candidate stars which are likely to be perturbed by a nearby dark subhalo, such that they could potentially be further analyzed with more detailed approaches. The search for dark subhalos in the reduced Gaia-like catalogs differs from the previous section on two fronts. On the one hand, the mock observed stars in the synthetic Gaia datasets are divided into patches using the hierarchical pixelization algorithm \textsc{HEALPY}~\citep{Zonca2019, 2005ApJ...622..759G} with a pixel level 6.
This allows to process the data in manageable subsets in a physically meaningful fashion.
In addition, as the halo-associated stars are located in well-defined, localized regions in the sky, we avoid using the absolute right ascension and declination coordinates to unfairly bias the model.
Instead, we compute the positional information with respect to the pixel center.
For a Gaia \acs{DR2}-like dataset, a pixel can contain up to $\simeq2\times10^{4}$ stars.

On the other hand, the input feature of each observed star is different. For each synthetic dataset realization $g \in \{{\tt m12f}, {\tt m12f}, {\tt m12m}\}$, $l \in \{{\tt LSR0}, {\tt LSR1}, {\tt LSR2}\}$, we then have a list of (star observation, label) pairs
$$
\mathsc{D}_{g,l} = [(\mathbf{X}_i, y_i), \dots].
$$
Each stellar feature vector $\mathbf{X}$ consists of the following astrometric observables:
\begin{itemize}
    \item parallax $p$ [mas],
    \item the right ascension with respect to the pixel center $\Delta \alpha$ [deg],
    \item the declination with respect to the pixel center $\Delta \delta$ [deg],
    \item the proper motion in the right ascension direction (multiplied by $\cos\delta$) $\mu_{\alpha}^*$ [mas/year],
    \item the proper motion in the declination $\mu_{\delta}$ [mas/year] and
    \item the radial velocity $v_r$ [km/s].
\end{itemize}
These observables are available with estimated uncertainties, resulting in 12 input features.
Features which are not always measured, such as the radial velocities, are filled with a placeholder value (numerically set to zero) for a consistent numerical treatment in the neural network model.

The anomaly detection model was trained for 200 epochs, while the classification model for 50 epochs.
As before, we use {\tt m12m} for the optimization, 
{\tt m12i} for the validation, while {\tt m12f} is used for testing.
The training and testing is done on stars from all three LSRs simultaneously.
Overall, as summarized in Table~\ref{tab:dataset}, the optimization, testing and final evaluation is carried out on nearly 1.5 billion mock stars, of which less than 0.01\% are identified as signal, resulting in extreme class imbalance as well as an overall low number of independent signal samples.

\begin{figure}[h]
    \centering
    \includegraphics[width=0.45\textwidth]{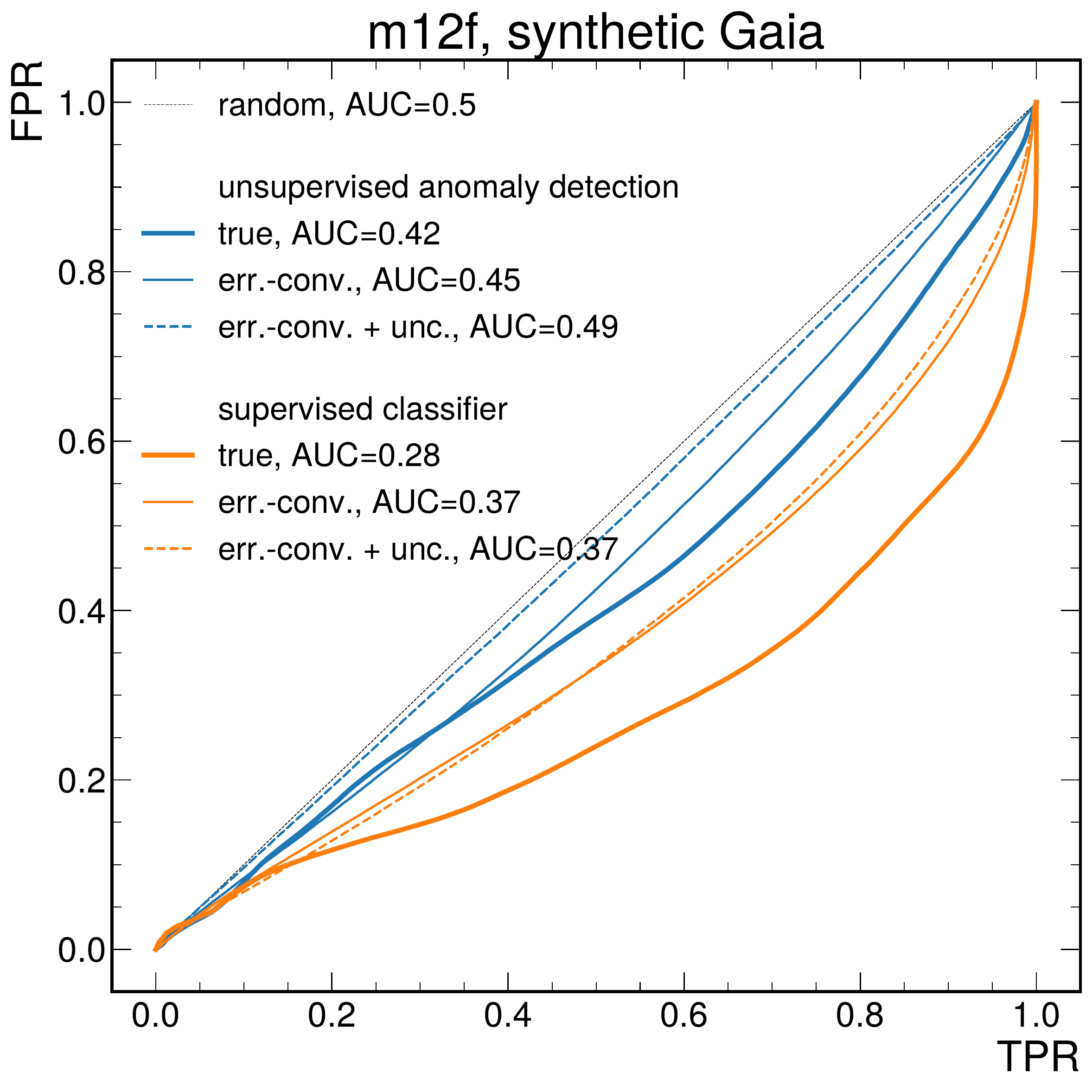}
    \caption{The true positive rate versus the false positive rate after evaluating the anomaly-detection (blue) and binary classification (orange) models on the synthetic Gaia dataset {\tt m12f}. We compare the model sensitivity with true inputs and error-convolved inputs with and without knowledge of the uncertainties.}
    \label{fig:gaia_roc_m12f}
\end{figure}

The sensitivity of the anomaly detection and classifier methods for identifying halo-associated stars in the synthetic Gaia dataset can be seen on Figure~\ref{fig:gaia_roc_m12f}.
As for the Latte runs, the {\tt m12f} dataset was never used in the optimization.
We observe that the binary classification distinguishes between the halo-associated and background stars at a non-negligible level, with a FPR of $\simeq35\%$ at a TPR of $\simeq50\%$.
On the other hand, the anomaly detection approach, where we only attempt to learn the background distribution, does not differ significantly from a purely random selection in the synthetic survey.

\section{Discussion}
\label{sec:discussion}

Based on the results presented in the last section, we conclude that the halo-associated star particles in the galaxy simulations adopted in this study have a distinguishable distribution in 6-dimensional phase-space (positions and momenta). The anomaly detection method is able to correctly identify halo-associated stars, where as the supervised binary classification does not perform well in the Latte simulations due to the very low signal statistics. When going from the idealized simulation to the synthetic survey, we observe, on the one hand, a mildly better performance for the supervised binary classifier (AUC $0.48 \rightarrow 0.37$). On the other hand, the unsupervised anomaly detection performs significantly worse (AUC $0.07 \rightarrow 0.45$).

It is important to note that even though the synthetic survey is derived from the Latte simulation, the results of both cases are not comparable, as multiple experimental and observational effects affect the synthetic dataset. In particular, there are the following two uncontrollable (at least for us) and random phenomena involved. First, the synthetic survey performance is limited by 
the sampling process that generates a population of synthetic stars from each star particle in the simulation. Thus introducing a smearing scale that might dilute the signal.
Second, you can get lucky (or unlucky) with the selection of LSRs. This introduces an observational bias since, by chance, more or less subhalos might be included in the footprint of the synthetic survey. On top of these, there is firstly the effect of simulated data uncertainties and secondly the fact that in each synthetic survey less than 0.5\% of the stars have measured radial velocities (see table~\ref{tab:dataset}), thus reducing the kinematic data from 6 to 5 dimensions. Finally, in the synthetic survey, the stellar disc was excluded in order to reduce the volume of data.

We studied the effect of simulated meausurement errors and the inclusion of radial velocities by redoing the synthetic Gaia analysis using the {\it true} astrometric inputs. As seen in figure~\ref{fig:gaia_roc_m12f}, the performance improves when changing from error-convolved to true values, but does not arrive to the one in the Latte simulation. Therefore, neither data uncertainties nor the absence of measured radial velocities have a major impact in driving the performance difference between the Latte simulation and the synthetic surveys. Furthermore, as also seen from figure~\ref{fig:gaia_roc_m12f}, providing information on the simulated uncertainties of the error-convolved values to the model does not significantly affect the supervised classification sensitivity, while it somewhat reduces the sensitivity of the autoencoder-based anomaly detection approach. The latter may be attributable to the increased difficulty of encoding 12 instead of 6 inputs.

Note that the name ``{\it true} inputs'' is misleading, since each synthetic star in phase-space has been sampled from a one-dimensional kernel centred on the generating star particle in position and velocity space~\cite{2020ApJS..246....6S}. 
We argue that the main difference in performance might be caused by this sampling process that introduces a smearing scale of the order of 0.7 kpc in position and roughly 10 km/s in velocity.\footnote{The smearing scales are defined as the standard deviation of the differences between the position or velocity of each halo-associated synthetic star and that of the parent star particle from which it was generated.} Since the change of a star's velocity due to the encounter with a subhalo of mass $M_{\rm sh}$ scales as $\sim 0.5-1\,{\rm km\,s^{-1}}\left(M_{\rm sh}/{10^{8}\,{\rm M_{\odot}}}\right)^{2/3}$~\cite{2015MNRAS.446.1000F} and kinematic perturbations are partially washed out below the smearing scale, the sampling process causes significant changes in the phase-space distribution of synthetic stars. It is therefore expected to dominate over the luck factor and the removal of the stellar disk in explaining the difference in performance between the idealized Latte simulation and the Gaia-like surveys.

Finally, we would like to highlight that to thoroughly investigate the above conclusion we need a set of dedicated simulations, where each possible effect can be turned on in sequence and can be easily disentangled. Given the scope of the additional studies required, we are studying this in a follow-up paper.

\section{Summary and Conclusions}
\label{sec:conclusions}
\acresetall

\Ac{ML} techniques, either alone or combined with classical methods, have been demonstrated to be helpful in uncovering new structures in Gaia-scale datasets (e.g. \citep{2020NatAs...4.1078N}).
Dark subhalos are among the most challenging substructures to search for.
In this paper, we study the detectability of dark subhalos by means of ML in three MW-like galaxies and in nine synthetic Gaia DR2 surveys. Rather than attempting to pinpoint the exact subhalo locations and determine their properties, we attempt to identify candidate stars that are likely to be close to a subhalo on a statistical basis.

We have first correlated star particles in the simulated galaxies and mock stars in the synthetic catalogs with the position of dark subhalos found by the Amiga halo Finder (AHF). 
In Section~\ref{subsec:detectability} we then tested the feasibility of an anomaly detection and a binary classification algorithm against simulated galaxies to detect the phase-space imprint in stellar halo stars of nearby subhalos. The first algorithm builds a likelihood function of the background star particles and is able to correctly identify 80\% of signal stars while misclassifying as signal 15\% of background particles. On the other hand, the binary classifier does not perform well due to the very low signal statistics. We concluded that the distribution function of the 6-dimensional phase-space coordinates of signal and background star particles are distinguishable in the MW-like galaxies used in this work. Therefore, on a statistical basis, position and velocity information can be combined into a statistical discriminator for the halo-associated signal. 

Finally, we have tested the feasibility of our algorithms in Gaia DR2-like surveys in Section~\ref{subsec:feasibility}. 
The anomaly detection approach has no sensitivity to distinguish between signal and background stars, while the binary classification algorithm is able to select 50\% of signal stars while wrongly identifying 15\% of background stars as signal. Although the binary classification shows a mild sensitivity, overall both approaches are of limited effectiveness in the synthetic Gaia survey. Although the results above are not directly comparable, our hypothesis is that the sampling process that generates a population of synthetic stars from each star particle mainly causes the difference in performance between the Latte simulation and the synthetic surveys. A thorough investigation of this conclusion is left to future work, leaving the use of ML-based tools as a new way to quantitatively study the effects of dynamical perturbations of DM subhalos as the main message of this paper.

A number of subsequent improvements to the methodology are possible. 
In the above analysis, all the observed stars were treated independently of each other.
Local correlations, density or clustering were not taken into account, which could potentially limit the sensitivity of the method used so far.
As an example, novel approaches based on density-based clustering have been employed for open clusters~\citep{castro2018new} and may be interesting to study here for dark subhalos.
Clustering can also be combined with unsupervised deep learning for anomaly detection~\citep{mikuni2021unsupervised}.
Another possible approach is the direct search for overdensities by comparing signal and sideband regions, which has been so far demonstrated for stellar streams, but could potentially be studied also for dark subhalos~\citep{10.1093/mnras/stab3372}.
Furthermore, in order to understand the potential sensitivity of the method, simulated datasets with a known \acs{DM} distribution were used.
However, the halo distribution in these is fixed, and the number of actual simulated halos in the potentially visible region is limited.
Additional simulated datasets with a varying halo distribution could be helpful to establish sensitivity dependence of a potential method on halo mass and distance from the galactic center.

Finally, Rubin/LSST will provide a deeper map of the Galactic stellar halo of the Milky Way compared to that of Gaia. 
For main-sequence stars, Rubin/LSST is expected to achieve a tangential velocity precision of $\mathcal{O}(10\,\rm km/s)$ up to Galactocentric distances of $20-30$ kpc, increasing up to $\sim 300$ km/s at distances of roughly $60$ kpc \citep{LSSTScience:2009jmu, LSST:2008ijt}.
For the MW-like galaxies adopted in our analysis, we find dark subhalos with masses $\sim 2\times 10^7\,\rm M_\odot$ ($\sim 2\times 10^8\,\rm M_\odot$) within 20-30 kpc (60 kpc) from the Galactic center. These subhalos induced velocity changes in the neighboring stars of $\sim 0.5$ km/s ($\sim 2$ km/s). Since these velocities changes are below the kinematic precision expected to be achieved with the Rubin/LSST, it will be a challenge to detect the kinematic effect of subhalos with this upcoming telescope.

\section*{Acknowledgments}
This work was supported by the Estonian Research Council grants \mbox{PSG700}, \mbox{PRG803}, \mbox{PRG1006}, \mbox{PUTJD907} and \mbox{MOBTT5}, \mbox{MOBTP187}, and by the European Regional Development Fund through the CoE program grant TK133. We would like to thank Martti Raidal for suggesting to us the idea of using large-scale ML approaches on the Gaia dataset, and for advice and support throughout this project. We also thank the referees for their time, effort and constructive input.

\bibliographystyle{model2-names}\biboptions{authoryear}



\end{document}